\documentclass[11pt,journal]{./IEEEtran}

\usepackage{graphicx}

\begin{document}

\title{Building Usage Profiles Using Deep Neural Nets
}

\author{\IEEEauthorblockN{Domenic Curro, Konstantinos G. Derpanis, Andriy V. Miranskyy \\}
\IEEEauthorblockA{Department of Computer Science, Ryerson University, Toronto, Canada\\
\{d2curro, kosta, avm\}@ryerson.ca
}
}

\maketitle

\begin{abstract}

 To improve software quality, one needs to build test scenarios resembling the usage of a software product in the field. This task is rendered challenging when a product's customer base is large and diverse.  In this scenario, existing profiling approaches, such as operational profiling, are difficult to apply.
In this work, we consider
publicly available video tutorials of a product to profile usage.
Our goal is to construct an automatic approach to extract information about user actions from instructional videos. To achieve this goal, we use a Deep Convolutional Neural Network (DCNN) to recognize user actions.
Our pilot study shows that a DCNN trained to recognize user actions in video can classify five different actions in a collection of 236 publicly available Microsoft Word tutorial videos (published on YouTube).  In our empirical evaluation we report a 
mean average precision of 94.42\% across all actions.
This study demonstrates the efficacy of DCNN-based methods for extracting software usage information from videos.  Moreover, this approach
may aid in other software engineering activities that require information about customer usage of a product.

\end{abstract}

\IEEEpeerreviewmaketitle

\section{Motivation}

As a software product's codebase grows, the number of execution paths grows exponentially \cite{jasper1994test}. This makes it impossible to test all conceivable execution paths, especially at the system-test level or above, where a tester must simulate how a client uses the product as a whole in practice \cite{jasper1994test, miranskyy2009selection}. 
Therefore, the tester needs to focus on the paths that users execute in the field.  This requires the tester to identify 
the paths clients actually use, as well as the popularity of those paths \cite{musa1993op}. To create a representative test scenario (i.e., workload covering execution paths traversed by the users in the field), the tester needs to know the action sequences that users perform along with their respective popularity among users. In this paper, \textit{usage profile} denotes the sequences of user actions and  popularity of these sequences among users.

A classic solution to this problem is operational profiling~\cite{musa1993op}.
To create a profile one needs to log information about execution paths covered by users (typically, actions can be extracted from the paths). 
Once the data are gathered, the popularity of a given path can be estimated based on its execution frequency.  In practice,
this approach is difficult to apply to every client. For instance, customers are reluctant to enable logging infrastructure on production systems, as it may lead to performance degradation, instability, and privacy breaches \cite{miranskyy2016operational}. Moreover, reaching every client (if the customer base is large) can be economically infeasible \cite{miranskyy2009selection}.

Another approach to identifying execution paths is by analyzing defects that users encounter and the sequences of actions/events needed to reproduce the defect \cite{miranskyy2009selection}. The popularity of a given execution path can be estimated by the number of encounters of this defect that users report to support personnel. A drawback of this approach is that it is biased 
towards problematic execution paths (as the paths target defect  reproduction). Moreover, not every defect encounter gets reported, e.g., because a user found a simple workaround or because the defect gets ``patched'' in the production codebase before a user executes a path containing a defect. 

\section{Goal and Potential}\label{sec:goal}

In this work, we consider non-traditional data sources to construct representative client usage profiles.  In particular, we use readily-available  
software product tutorials posted on video-sharing websites, such as YouTube.
The tutorials allow an analyst to reconstruct the sequence of user actions to achieve their goal. The analyst can then assess the popularity of this sequence by looking at the number of views of a given video, as well as its rating (the views and rating data are publicly available). One can assume that the higher the number of views and rating of a video, the higher the probability that a given sequence will be used by clients in the field. Thus, the analyst can obtain two pieces of data needed to construct a usage profile: a sequence of actions and its popularity.

Information about the number of views and rating of the videos can be easily obtained using YouTube's API.  In contrast,  manually obtaining the action sequences described in the videos is prohibitively expensive due the larger number of videos
and their significant lengths.
For example, at the time of writing, there were 
approximately $286,000$ videos (based on the number of videos found on youtube.com for the search keywords ``microsoft word tutorial'').  This motivates the need for automatic means to  
extract information about actions and their sequences from videos.

\section{Method}\label{sec:method}

Deep Convolutional Neural Networks (DCNNs) \cite{jarrett2009,krizhevsky2012imagenet} have emerged as the standard approach for image understanding tasks, e.g., object recognition \cite{imagenet}. 
Recent work has demonstrated that the features learned in a DCNN are transferable to other related tasks~\cite{sharif2014cnn,harley2015document}.  As a result,
this reduces the amount of data required to train a network.  In particular,
rather than train a network from scratch, one begins with a pretrained
network and fine-tunes the network's parameters based on
their (possibly limited) training data for the task at hand. 
Leveraging these advancements, we demonstrate that videos posted on websites, such as YouTube, are a rich source of untapped user-profiling data.

We constructed our dataset (available at~\cite{curro_domenic_2017_321921})  by extracting the salient video frames.  Frames were deemed salient when they were distinct enough from their neighbouring frames by a simple image difference approach.
This ensures a wide variety in image appearances. The images were resized to $256 \times 256$ pixels, and labelled based on the apparent user action.

\begin{figure*}[t]
    \centering
    \includegraphics[width=\textwidth]{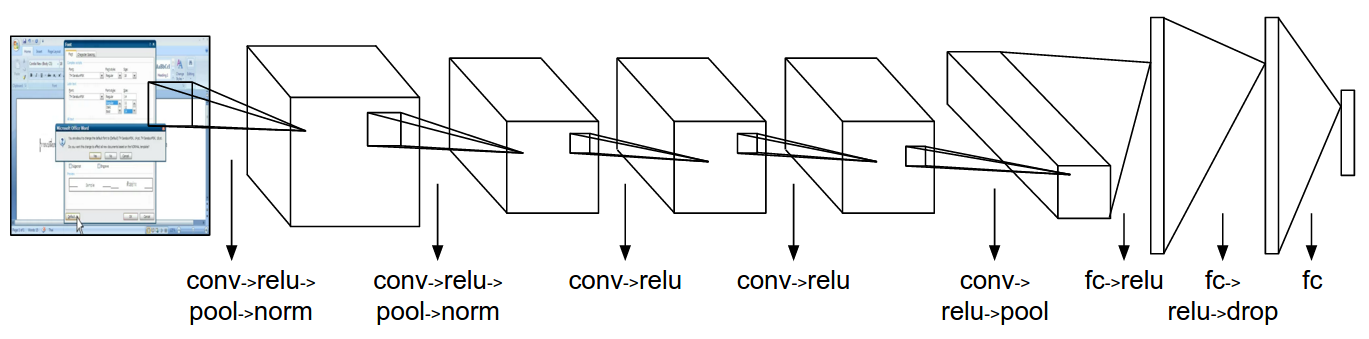}
    \caption{ Our DCNN architecture,  based on Caffe~\cite{jia2014caffe} version of AlexNet~\cite{krizhevsky2012imagenet}. }
    \label{fig:alexnet_arch}
\end{figure*}

The DCNN model was trained with Caffe~\cite{jia2014caffe}, a popular open source deep learning framework.  There are a variety of DCNN architectures used
for image understanding.  In this work, we used the 
AlexNet architecture~\cite{krizhevsky2012imagenet}, a standard DCNN baseline, summarized in Figure~\ref{fig:alexnet_arch}, 
which is a formulation of a Neural Network that is designed for image processing. 
It consists of five convolutional layers, which perform a discrete sliding window style image filtering with each element of the filter being a learnable weight; three max-pooling layers, which perform downsampling; three fully-connected layers, which compute the inner products of the learnable weights and the input feature vector; one dropout layer which deactivates 50 percent of the units randomly, adding a form of regularization; and a softmax loss layer, which performs a normalized multinomial logistic function of the output of the final fully-connected layer, producing class probability confidences. Each convolutional and fully-connected layer, except the final two, are followed by a non-linear rectified linear unit (ReLU) layer. 

 We leveraged knowledge transfer (in the spirit of \cite{sharif2014cnn, harley2015document}, as discussed above) by initializing the convolutional layers with the parameters learned by AlexNet for the ImageNet challenge~\cite{imagenet}: a large scale natural image classification challenge consisting of 1000 classes~\cite{imagenet} ranging from different species of dogs, to buildings and structures. The model was trained for three epochs\footnote{An epoch is the number of iterations to process the entire training set.}, using Stochastic Gradient Descent with a momentum of 0.9, weight decay of 0.0005, and a batch size of 128 images. The learning rate was initialized to 0.001 and was annealed by a factor of 0.1 every epoch. The network's training error is computed with a multinomial logistic loss~\cite{krizhevsky2012imagenet}. The input images are zero centered by subtracting the pre-calculated mean image from the ImageNet dataset. During training, we performed small random translations and randomly crop each image to $227\times 227$ pixels. This step artificially increases the training set, allowing for some spatial invariance~\cite{simard2003document,ciregan2012computer}. At test time, only the center $227 \times 227$ pixels of the image are evaluated.

To profile user behaviour, each video was manually labelled with the class of the sequence of actions occurring in the video. When a video contains more than one class, it was split up into video clips, with one class per clip (standard practice within the action recognition community~\cite{iccv13-action}).

Profiling is achieved by first classifying each frame of our test video to generate a time series of softmax-confidence class scores. Essentially, given $n$ frames in a video, we get a sequence of characters, $s_1, s_2, \ldots, s_n$, where $s_i$ represents the class of the $i$-th frame. Regular expressions are used to localize the desired user action, within the sequence. (More sophisticated
approaches to reasoning about action sequences are possible and reserved for future work.) The expression that returns the highest confidence score is considered the predicted execution path. In the case that no expressions matched~\cite{perlre}, the frame with the highest confidence is mapped to its highest related execution path.

Performance is measured by analyzing a confusion matrix\footnote{Each row and column sums to the image count per class, and predictions per class, respectively. }, precision, recall, and F1-score\footnote{The weighted average of the precision and recall.}. To obtain overall model performance for classification of the video clips, we compute the Average Precision $AP$, a standard measure of model performance in computer vision\footnote{$AP$ is more sensitive than the Area Under the Curve measure~\cite[Sec. 4.2]{everingham2010pascal}.}:
$AP = 1/11 \sum_{r \in \{0.0, 0.1, \ldots, 1.0 \}} { \max_{ \tilde{r} : \tilde{r} \ge r } p(\tilde{r}) }$,
where $p(\tilde{r})$ is the measured precision at recall $\tilde{r}$; see~\cite[Sec. 4.2]{everingham2010pascal} for details.
All measures range between 0 and 1; the higher the value -- the better the performance.

\begin{table*}[ht]
\centering
\caption{10-fold cross validation confusion matrix -- classification of individual images, the legend is given in Section~\ref{sec:act}.}
\label{tbl:confusion_img}
\begin{tabular}{r|rrrrrr|r}
& $b$   & $f$   & $F$   & $c$   & $C$   & $p$   & \%recall \\
\hline
$b$ & 38852 & 49    & 25    & 210   & 170   & 198   & 98.35      \\
$f$ & 27    & 86    & 2     & 0     & 3     & 0     & 72.88      \\
$F$ & 8     & 4     & 33    & 0     & 1     & 0     & 71.74      \\
$c$ & 34    & 0     & 0     & 347   & 0     & 0     & 91.08      \\
$C$ & 93    & 0     & 1     & 2     & 300   & 0     & 75.76      \\
$p$ & 39    & 0     & 0     & 0     & 1     & 253   & 86.35      \\
\hline
& 99.49 & 61.87 & 54.10  & 62.08 & 63.16 & 56.10  & \multicolumn{1}{l}{\%precision}   \\
& 98.91 & 66.93 & 61.68 & 73.83 & 68.89 & 68.01 & \multicolumn{1}{l}{\%F1-score}
\end{tabular}
\end{table*}

\section{Pilot Study}

\subsection{Data Preparation}

We manually assembled a dataset consisting of 236 Microsoft Word video clips, downloaded from YouTube. The dataset contains tutorial videos of users explaining how to achieve the following three goals: (i) change the default font, (ii) choose the number of columns in the document, and (iii) add page numbers to the document, for Word 2007, 2010, and 2013.

\begin{figure*}[t]
    \centering
    \includegraphics[width=\textwidth]{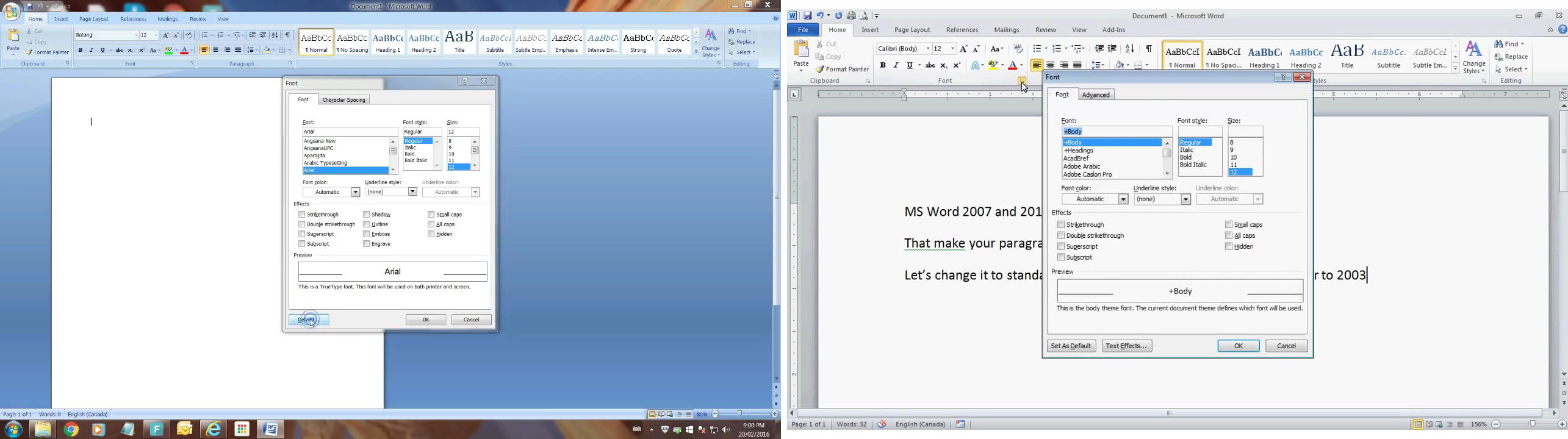}
    \caption{ Font window variability in two different video clips.}
    \label{fig:var_example}
\end{figure*}

 The resolutions of the videos range between $294 \times 240$ and $1920\times 1080$.  The videos contain variety of user interface (UI) changes, screen-capturing software artifacts, intro and outro segments, screen tones and colors, user themes, tutorial artifacts, mouse occlusions, and varying system fonts.  Example frames from the dataset are 
shown in Figure~\ref{fig:var_example}.

\noindent{\bf Individual actions}\label{sec:act} Our dataset consists of a subset of frames from the original videos.  Each frame was individually labelled based on its dominant class: 
$f$, Font window; 
$F$, Default Font window; 
$c$, Column dropdown;
$C$, Column window; 
$p$, Page Number; and
$b$, Background.
There are 118 $f$, 46 $F$, 381 $c$, 396 $C$, 293 $p$, and 39,504 $b$ frames.

\noindent{\bf Sequences of actions}\label{sec:seq} Each individual video clip was labelled to indicate the apparent user execution path. A video clip contains one the following five sequences of user actions: 
$\alpha$, opening the Font window, followed by opening the Default Font window; 
$\beta$, opening the Font window, but not opening the Default Font window; 
$\gamma$, opening the Column Dropdown menu, selecting More Columns, opening the Columns Window; 
$\delta$, opening the Column Dropdown menu, but not selecting More Columns; 
$\epsilon$, selecting the Page Number Dropdown. 
There are 6 $\alpha$, 17 $\beta$, 80 $\gamma$, 58 $\delta$, and 75 $\epsilon$ clips. Note that $\alpha$ and $\beta$, and $\gamma$ and $\delta$ represent mutually exclusive execution paths dedicated to fulfilling the same goal: setting the font and setting the number of columns, respectively. This shows that there can be multiple execution paths used to achieve a relatively simple goal.

\subsection{Classifying Individual Images / Actions}\label{sec:res_img}

We used 10-fold cross validation to train and evaluate our approach. The images were split into 10 subsets, where each fold contains a unique permutation of eight subsets as the training set, one as the validation set, and one as the test set.

 We trained one model per fold, monitoring its performance on the respective validation subset. 
When the validation loss plateaued, training was manually stopped. Visualizing a loss (lower is better) over iteration curve, candidate iterations were chosen and their F1-scores and confusion matrices were produced. The best iterations ranged between 2133 and 4085 (with 3270 iterations on average), indicating that the best model is not always found at the final iteration of training. The average logistic loss for the best model of each fold ranged between 0.069 to 0.478  (0.168 averaged over all of the folds), suggesting that some models  perform better than others.

Evaluating our models, we classify the images in each fold's test set. The performance of our approach is summarized in Table~\ref{tbl:confusion_img}, which depicts an amalgamated confusion matrix of all ten folds (covering 40,738 images), as well as the precision, recall, and F1-scores. The background class, having the most examples, scored the highest on the  F1-score  at 98.91\%. The remaining F1-scores range between 73.83\% to 61.68\%.

\subsection{Classifying Videos / Sequences of Actions}\label{sec:res_seq}

Each frame of a video clip is associate with one or more user actions. Classifying the sequence of frames of a video, we generate a sequence of actions, allowing us to profile the customer's use of the product (discussed in Section~\ref{sec:goal}). 

First, the trained DCNN model\footnote{Models only predict video clips that were used to compose their test set.} is used to predict the class of each frame of a test set video.  This generates a sequence of class prediction softmax-confidence scores, as discussed in Section~\ref{sec:method}.
The sequences are then average-smoothed using a one second wide kernel (sliding window style smoothing). We also tried using a Gaussian kernel, but average-smoothing provided the best results.
Then, to determine the user behaviour (also discussed in Section~\ref{sec:method}), we applied regular expressions to these sequences of characters as follows.

Discovering if the user: 

$\alpha$) sets their font via the Font window, we search for the font menu appearing for at least one second: \texttt{f\{r,\}}\footnote{In regular expression notation \texttt{x\{a,b\}} represents \texttt{x} occurring \texttt{a} to \texttt{b} times; \texttt{x\{a,\}} -- \texttt{x}
occurring \texttt{a} or more times.}, where \texttt{r} is the frame rate of the video, which acts as one second of video time in this context.

$\beta$) sets their default font, we search for the Font window appearing, followed by the Default Font window appearing: \texttt{f\{r,\}F\{r,\}f\{0,r\}}. When the default font prompt is closed, for a short while, the font menu may remain open.

$\gamma$) sets the number of columns in the document via the Column dropdown menu, we search for the Column dropdown menu opening:  \texttt{c\{r,\}}.

$\delta$) sets the number of columns in the document via the Column window, we search for the Column dropdown opening, followed by the Column window appearing: \texttt{c\{r,\}[\string^cC]\{0,r\}C\{r,\}}. We allow time between the column drop down menu closing, and for the columns pop-up window finally appearing, accounting for noisy intermediate predictions of neither \texttt{C} nor \texttt{c}.

$\epsilon$) sets the page number, we search for the Page Number dropdown menu opening:  \texttt{p\{r,\}}.

Mandating that one second of a desired classification occurs eliminates one-off random false positive predictions, caused by the individual frame classification process. 

When a regular expression is matched, that region of the sequence is removed to prevent any related regular expression from considering it. Thus, the user setting her default font via the Default Font window ($\beta$) must be searched for and removed prior to the user setting her font via the Font window ($\alpha$). Similarly, ($\delta$) should be searched for before ($\gamma$).

Results are summarized in Table~\ref{tbl:confusion_vid}.  Overall, the sequence classifier achieves a mean Average Precision of 94.42\% (computed using Pascal VOC challenge toolbox); the $AP$ ranges between 80.16\% and 99.82\%.

\begin{table*}[ht]
\centering
\caption{10-fold cross validation confusion matrix -- classification of the videos, the legend is given in Section~\ref{sec:seq}. }
\label{tbl:confusion_vid}
\begin{tabular}{r|rrrrr|r}
& $\alpha$     & $\beta$     & $\gamma$     & $\delta$     & $\epsilon$     & \%recall                   \\
\hline
$\alpha$ & 4     & 1     & 1     & 0     & 0     & 66.67                        \\
$\beta$ & 2     & 15    & 0     & 0     & 0     & 88.24                        \\
$\gamma$ & 1     & 0     & 75    & 2     & 2     & 93.75                        \\
$\delta$ & 0     & 0     & 4     & 54    & 0     & 93.10                        \\
$\epsilon$ & 0     & 0     & 1     & 0     & 74    & 98.67                        \\
\hline
& 57.14 & 93.75 & 92.59 & 96.43 & 97.37 & \multicolumn{1}{l}{\%precision} \\
& 61.54 & 90.91 & 93.17 & 94.74 & 98.01 & \multicolumn{1}{l}{\%F1-score}     \\
& 80.16 & 95.1 & 97.25 & 99.82 & 99.79 & \multicolumn{1}{l}{\%$AP$}    
\end{tabular}    
\end{table*}

\section{Discussion and limitations}

Training was performed using an nVidia Titan X Pascal GPU card.  Training takes
approximately thirty minutes. Validation speed was approximately 530 images per second. Thus, one hour of GPU time could potentially process 6.8 hours of YouTube video (assuming that we sample 1 frame per second), making it applicable for practical applications. 
Since the videos are independent of each other, processing can be easily parallelized on multiple GPU cards.

The original AlexNet DCNN was trained on natural images~\cite{krizhevsky2012imagenet} (e.g., pictures of dogs, cats, and buildings), while we  deal with a significantly different class of images from the UI domain. Nevertheless, our models have good predictive power, suggesting that at some level of abstraction (e.g., shape and thickness of the lines in the image) the key factors differentiating classes of natural and UI images are similar. Borrowing pretrained features from a network trained in a related domain has been demonstrated to yield better results \cite{sharif2014cnn, harley2015document}. Thus, initializing our network with the learned features from a network specialized on UI images would likely yield higher classification power.

Note that the performance of the video clip classifier (discussed in Section~\ref{sec:res_seq}) is higher than that of the classifier of a single image  (discussed in Section~\ref{sec:res_img}). This can be explained by the increase of the amount of information in the sequence of frames in comparison with a single image: e.g., if we misclassify one frame in a sequence of ten frames, we will still be able to use the correct information from the remaining nine frames.

\noindent{\bf Limitations} There are two major drawbacks with the proposed approach.  First,
DCNNs require a large volume of images for training to achieve high predictive power~\cite{krizhevsky2012imagenet}. As a result, this approach may not be applicable to actions containing a relatively small number of training videos.  Second,
GPUs outperform CPUs in training and validating DCNNs by one to two orders of magnitude~\cite{shi2016benchmarking}. Thus, specialized hardware is required 
to speedup computation.

\section{Related work}
Video recordings are already used in Software Engineering (SE); e.g., in ethnographic studies of development organizations and in user experience research (see \cite{socha2016wide} for review). To the best of our knowledge, automatic classification of sequences in videos is not utilized. 

There also exist tools for automatic user interface testing that search for a specific image on the screen~\cite{yeh2009sikuli}. However, these tools can only match user provided images to areas on the screen. While they do provide a threshold-error to allow near matches, they cannot account for unpredictable variation on the screen. While both approaches can automatically profile user action, only ours remains invariant to scale, color change, and version style change; thus, reducing the amount of manual labour and saving analysts' time.

It was suggested~\cite{white2015deep} that DCNNs may be used in source code analysis for ``... viz. code suggestion, code summarization, traceability link recovery, and feature location''. To the best of our knowledge, no-one in SE has considered extracting UI-based actions from videos using DCNN.

\section{Conclusions and Future Work}

In this work, we have studied the applicability of DCNNs to the extraction of user activities from video tutorials, showing that the extraction of information about user activities  from video clips can be automated. Moreover, DCNNs are capable of generalizing to multiple versions of the UI. The information about the activities may be utilized in building usage profiles (e.g., to construct representative test scenarios/workloads) or in other software engineering activities that can leverage information about usage of a product.
The approach is fast (processing approximately seven hours of video per GPU-hour) and scalable (as it can be easy parallelized on a GPU cluster). The predictive power of the DCNN model is high:  the mean average precision is 94.42\% on the five action sequences
considered in our dataset~\cite{curro_domenic_2017_321921} comprised of 236 Microsoft Word tutorial videos (published on youtube.com).

Going forward, we plan to expand the study to other products and features, explore statistical techniques for classifying sequences of images, and develop custom-built DCNNs tailored for the UI domain.

\bibliographystyle{abbrv}
\bibliography{ref}

\end{document}